\documentclass[aps,prl,twocolumn,superscriptaddress,showpacs,amsmath,amssymb,floatfix,nofootinbib]{revtex4-2}
\usepackage{braket}
\usepackage{natbib}

\begin{document}

\title{\Large Two pseudo-Goldstone bosons in the Gildener-Weinberg model}

\author{Parsa Ghorbani}
%\email{parsaghorbani@gmail.com}
\affiliation{Physics Department, Faculty of Science, Ferdowsi University of Mashhad, Iran}
%\author{Second Author}
%\email{second.author@example.com}
%\affiliation{Department of Physics, University Y, City, Country}

%\date{\today}

\begin{abstract}
In a dimensionless multi-scalar extension of the Standard Model, Gildener and Weinberg assumed that along a flat direction, there is only one classically massless scalar, known as the {\it scalon}, which acquires mass through radiative corrections à la Coleman-Weinberg, while all other scalars remain heavy. In this paper, by introducing a toy model with four scalar degrees of freedom we demonstrate the existence of {\it two} scalons along a specific flat direction that we construct. We present the effective potential for the model and provide the masses of the heavy scalars and two radiatively light pseudo-Goldstone bosons.
\end{abstract}


\maketitle

\section{Introduction}
Gildener and Weinberg (GW) \cite{Gildener:1976ih} came up with a generic dimensionless multi-scalar model as an extension to the Standard Model Higgs potential for which the theory undergoes a dynamical electroweak symmetry breaking through the radiative corrections  \`a la Coleman-Weinberg \cite{Coleman:1973jx}. 

In the approach of GW, a quartic multi-scalar potential $V=\lambda_{ijkl}\phi_{i}\phi_{j}\phi_{k}\phi_{l}$ can be represented in {\it radial form} as $V=\phi^4 \lambda_{ijkl} N_i N_j N_k N_l$, if we replace the scalar fields $\phi_i$ by $\phi_i=\phi N_i$ with $\phi \in (0,\infty)$ and $N_i$ being respectively the radial field, and a unit vector as a ray in the field space from origin to a point on the surface of the unit sphere. Then at a given mass scale $\Lambda$, a flat direction ${\bf N=n}$ for which ${\bf n}.{\bf n}=1$, is found such that  the potential and its minimum vanish; i.e. $\lambda_{ijkl} n_i n_j n_k n_l=0$ and $\lambda_{ijkl} n_j n_k n_l=0$. According to Coleman-Weinberg mechanism, loop quantum corrections to the effective potential lead to the generation of non-zero vacuum expectation value (VEV) for the radial field $\phi$ giving VEV to all scalars through $\braket{\phi_i}=\braket{\phi} n_i$. Whether or not  a scalar field in the model takes a VEV, depends on the choice of the flat direction. This non-zero $\braket{\phi}$ breaks symmetries of the underlying theory.

In their seminal paper \cite{Gildener:1976ih}, Gildener and Weinberg assume, without providing any proof, that at tree level there is only one massless scalar (regardless of the total number of scalars) due to the scale symmetry breaking. They name this scalar the "scalon" which gains a slight mass due to radiative corrections. All other scalars remain massive. 
Note that any other spontaneously broken symmetry which is space-time independent, would result in additional massless Goldstone bosons. Subsequently, the mass matrix $P_{ij}=1/2 \lambda_{ijkl} n_k n_l$ (see Eq. (3.8) in \cite{Gildener:1976ih}), will have more than one zero eigenvalues. \footnote{An example is the global $U(1)$  symmetry in complex singlet scalar extension of the SM, see e.g. \cite{Alexander-Nunneley:2010tyr}.}. Here, we are not considering any global symmetry of this type.

In this paper we question the uniqueness of the scalon in scale invariant multi-scalar potentials. Note that applying the Goldstone theorem for spontaneously broken space-time dependent symmetries including the scale invariance, is not straightforward \cite{Coradeschi:2013gda}. The counting rule, $n_{GB}=\text{dim} G -\text{dim} H$ with 
$n_{GB}$,  dim $G$ and dim $H$ being respectively, the number of Goldstone bosons, the number of generators before symmetry breaking, and the number of generators of unbroken symmetry, does not always hold \cite{Nielsen:1975hm,Low:2001bw, Coradeschi:2013gda}. As discussed in \cite{Nielsen:1975hm}, the number of Goldston bosons in a spontaneous symmetry breaking is equal to or greater than the number of generators in broken symmetry. 
In this work, we are not going to provide a formal proof of the number of Goldstone bosons for spontaneous scale symmetry breaking. Instead, we aim to show that in the absence of any space-time independent global symmetry in the model we will provide in four-dimensional space-time, there exist two Goldstone bosons. We emphasize that these massless modes are not related to any space-time indepoendent global symmetry such as global $U(1)$ symmetry. 

We will argue that despite the GW approach, it is not necessary to be restricted to only one radial field or flat direction. In fact, instead of having a single flat direction $n_i$ with $n_i n_i=1$, we will identify two flat directions, $n_j$ and $l_k$, such that $n_j n_j + l_k l_k=1$. Essentially, this implies that instead of replacing $\phi_i$ by $\phi N_i$ for all scalar fields in the potential, we choose two radial fields, say $\rho$ and $\eta$. For a subset of scalars, we have $\phi_j=\rho N_j$, and for the remaining ones, $\phi_k=\eta L_k$, where $N_j$ and $L_k$ represent two arbitrary directions in the field space. Having defined two flat direction in terms of the radial fields $\rho$ and $\eta$, the potential takes the general form as follows
\begin{equation}
\begin{split}
V=&\rho^4 \lambda_{ijkl} N_i N_j N_k N_l+\eta^4 \lambda'_{ijkl} L_i L_j L_k L_l\\
&+\rho^2 \eta^2 \lambda''_{ijkl}  N_i N_j L_k L_l \,.
\end{split}
\end{equation}
Note that the original GW conditions must still be satisfied; that is along both flat directions one must have
\begin{equation}\label{vcon}
\begin{split}
&V_\text{tr}({\bf N, L}) \bigg|_{\tiny\begin{matrix} {\bf N=n}  \\ {\bf L=l} \end{matrix}}=0\\
&   \frac{\partial V_\text{tr}({\bf N, L}) }{\partial N_i} \bigg|_{\tiny\begin{matrix} {\bf N=n} \\ {\bf L=l} \end{matrix}}
  = \frac{\partial V_\text{tr}({\bf N, L}) }{\partial L_i} \bigg|_{\tiny\begin{matrix} {\bf N=n} \\ {\bf L=l} \end{matrix}} 
 =0 .
 \end{split}
\end{equation} 
where $\bf N=\bf n$ and $\bf L=\bf l$ indicate two flat directions.
\section{Gildener-Weinberg approach}
To employ the GW approach with one scalon, let us examine a straightforward $\mathbb{Z}_2$ symmetric potential featuring two real scalars. The tree-level potential is given by 
\begin{equation}\label{v2scal}
 V(h_1, h_2)=\frac{1}{4}\lambda_{1}h_{1}^{4}+\frac{1}{2}\lambda_{12} h_{1}^{2}h_{2}^{2}+\frac{1}{4}\lambda_{2}h_{2}^{4}\,.
\end{equation}
This structure could represent a scale-invariant  real singlet scalar extension to the Higgs potential
\begin{equation}
V(H_1, h_2)=\lambda_{1}\left(H_1^\dagger H_1 \right)^2+\lambda_{12} \left(H_1^\dagger H_1 \right) h_{2}^{2}+\frac{1}{4}\lambda_{2}h_{2}^{4}\,,
\end{equation}
where $H_1^\dagger = (0, h_1)/\sqrt{2}$ denotes the Higgs doublet in the unitary gauge. However, our intention is not to complicate the model with internal symmetries that are broken after the fields acquire VEV. Instead, we focus solely on the potential presented in Eq. (\ref{v2scal}) involving two singlet scalars.  The flat direction $(n_1, n_2)$ is determined by the conditions 
\begin{equation}\label{fc1}
\lambda_{12}^{2}=\lambda_{1}\lambda_{2},~~~ \frac{\braket{h_1}^2}{\braket{h_2}^2}\equiv\frac{n_1^2}{n_2^2}=-\frac{\lambda_{12}}{\lambda_1},~~~ \lambda_{12}<0.
\end{equation}
with $n_1^2+n_2^2=1$. This is equivalent to having
\begin{subequations}\label{fc2}
\begin{align}
 \lambda_1 =-\lambda_{12} \frac{\braket{h_2}^2}{\braket{h_1}^2}\,,\\
 \lambda_2 =-\lambda_{12} \frac{\braket{h_1}^2}{\braket{h_2}^2}\,.
 \end{align}
\end{subequations}
with $\lambda_{12}<0$. Note that the condition $\lambda_{12} < 0$ is needed, along with $\lambda_{1} > 0$ and $\lambda_{2} > 0$, for the potential to be bounded from below along the flat direction.
The potential in terms of a single radial field is
\begin{equation}\label{vrad}
 V_\text{tr}({\bf N})=  \rho^4 \left( \frac{1}{4} \lambda_1 N_1^4+\frac{1}{4}\lambda_2 N_2^4+\frac{1}{2} \lambda_{12} N_1^2 N_2^2 \right).
\end{equation}
where ${\bf N}=(N_1,N_2)$ is an arbitrary direction in the field space.
Along the flat direction given by Eq. (\ref{fc1}) or Eq. (\ref{fc2}), the tree-level potential in Eq. (\ref{vrad}) satisfies the conditions
\begin{equation}\label{1scaloncon}
V_\text{tr}({\bf N}) \bigg|_{{\bf N=n} }
  = \frac{\partial V_\text{tr}({\bf N})}{\partial N_i} \bigg|_{{\bf N=n}} 
 =0 \,.
\end{equation} 
The mass matrix from the tree level potential is not diagonal; the scalar masses are the eigenvalues of the mass matrix as 
\begin{equation}\label{m1m2}
 m_1^2=0,~~~ m_2^2= -2  \lambda_{12} \left(\braket{h_1}^2 +\braket{h_2}^2 \right)\,.
\end{equation}
As anticipated by the GW scheme, there exists one massless scalar or scalon at tree level, while the other state is massive. 
The corresponding one-loop effective potential, using dimensional regularization in the $\overline{\text{MS}}$ scheme, is given by \cite{Alexander-Nunneley:2010tyr, Ghorbani:2015xvz},
\begin{equation}
V_\text{eff}({\bf n})\equiv V_\text{tr}({\bf n})+V_\text{1-loop}({\bf n})=B({\bf n})\rho^4 \left( \log \frac{\rho^2}{\braket{\rho}^2}-\frac{1}{2}\right),
\end{equation}
with
\begin{equation}
B({\bf n})=\frac{m_2^4}{64 \pi^2 \braket{\rho}^4}.
\end{equation}
This leads to radiative correction to the scalon mass as
\begin{equation}
\delta m_1^2=\frac{m_2^4}{8\pi^2 \braket{\rho}^2},
\end{equation}
where $m_2^2$ is known from Eq. (\ref{m1m2}).

\section{ Two-scalon approach}
Finding the flat direction with one scalon is simple for a potential with two real scalars, but it becomes increasingly complicated as the number of scalars grows \cite{Kannike:2019upf}. A relevant question is whether we can discover two flat directions corresponding to two scalons, giving rise to two pseudo-Goldstone bosons due to quantum corrections. To address this question, we examine a toy model featuring four real singlet scalar degrees of freedom,
\begin{equation}\label{Vtr}
\begin{split}
 V_\text{tr}=&\lambda_{0} h_1 h_2 h_3 h_4 +\frac{1}{4} \lambda_1 h_1^4+\frac{1}{4}\lambda_2 h_2^4+\frac{1}{4}\lambda_3 h_3^4+\frac{1}{4}\lambda_4 h_4^4\\
 &+\frac{1}{4} \lambda_{12} h_1^2 h_2^2+\frac{1}{4} \lambda_{13} h_1^2 h_3^2+\frac{1}{4} \lambda_{14} h_1^2 h_4^2\\
 &+\frac{1}{4} \lambda_{23} h_2^2 h_3^2+\frac{1}{4} \lambda_{24} h_2^2 h_4^2+ \frac{1}{4} \lambda_{34} h_3^2 h_4^2,
 \end{split}
\end{equation}
with the discrete $\mathbb{Z}_2\times \mathbb{Z}_2\times \mathbb{Z}_2$ symmetry under which the scalar fields transform as
\begin{equation}\label{z2z2}
\begin{split}
\mathbb{Z}_2:~ h_1\to -h_1, ~~~ h_2\to h_2, ~~~ h_3\to -h_3,~~~h_4\to h_4,\\
\mathbb{Z}_2:~ h_1\to -h_1, ~~~ h_2\to -h_2, ~~~ h_3\to h_3,~~~h_4\to h_4,\\
\mathbb{Z}_2:~ h_1\to h_1, ~~~ h_2\to h_2, ~~~ h_3\to -h_3,~~~h_4\to -h_4.
\end{split}
\end{equation}
Note the presence of an ad-hoc {\it quad-linear} term in the potential. The linear and cubic terms, as well as  terms of the form $h_i h^3_j$ for any $i,j=1,2,3,4$ are absent because of the $\mathbb{Z}_2\times \mathbb{Z}_2 \times \mathbb{Z}_2$ symmetry. 

Again, it should be noted that our focus here is not on  internal symmetries of the underlying theory. However, in order to relate this model to a more phenomenological one beyond the Standard Model framework, consider a potential consisting of two Higgs doublets and two real singlet scalars (For scale-invariant scenarios of 2HDM, refer to \cite{Lee:2012jn, Plascencia:2015xwa, Karam:2015jta})
\begin{equation}
\begin{split}
V_\text{tr}= &\lambda_0 (H_1^\dagger H_2 +H_2^\dagger H_1) h_3 h_4\\
&+ \lambda_1 (H_1^\dagger H_1)^2
+\lambda_2 (H_2^\dagger H_2)^2\\
&+\lambda_{12}' (H_1^\dagger H_1)(H_2^\dagger H_2)
+ \lambda_{12}'' (H_1^\dagger H_2)(H_2^\dagger H_1)\\
&+\lambda_{12}'''(H_1^\dagger H_2 + H_2^\dagger H_1)^2 
+\frac{1}{2}\lambda_{13} (H_1^\dagger H_1) h_3^2\\
&+ \frac{1}{2}\lambda_{14} (H_1^\dagger H_1) h_4^2
+ \frac{1}{2}\lambda_{23} (H_2^\dagger H_2) h_3^2\\
&+ \frac{1}{2}\lambda_{24} (H_2^\dagger H_2) h_4^2 +\frac{1}{4} \lambda_{34}  h_3^2 h_4^2+\frac{1}{4}\lambda_3 h_3^4
+\frac{1}{4}\lambda_4 h_4^4,
\end{split}
\end{equation}
where all the couplings are real. In this context, where $H_1$ and $H_2$ represent the doublets, and $h_3$ and $h_4$ denote the singlets, the discrete symmetry outlined in Eq. (\ref{z2z2}) manifests as a mechanism to prevent Flavor-Changing Neutral Currents (FCNCs). 

 If we were to follow the GW prescription in search of a flat direction $(n_1, n_2, n_3, n_4)$ that satisfies the conditions in Eq. (\ref{1scaloncon}), the solution would be extremely challenging, if not unattainable. Instead, following the approach outlined above, at a given scale $\Lambda$ we define two flat directions $(n_1, n_2)$ and $(l_1, l_2)$ that satisfy Eq. (\ref{vcon}). The result is the conditions 
 
 \begin{subequations}
    \noindent\centering
    \begin{minipage}{0.48\textwidth}
        \begin{align}
        \lambda_{23}=-\lambda_{0} \frac{\braket{h_1} \braket{h_4}}{\braket{h_2} \braket{h_3}}\\
 \lambda_{13}=-\lambda_{0} \frac{\braket{h_2} \braket{h_4}}{\braket{h_1} \braket{h_3}}\\
 \lambda_{14}=-\lambda_{0} \frac{\braket{h_2} \braket{h_3}}{\braket{h_1}\braket{ h_4}}\\
 \lambda_{24}=-\lambda_{0} \frac{\braket{h_1} \braket{h_3}}{\braket{h_2} \braket{h_4}}
        \end{align}
    \end{minipage}
    \hfill
    \begin{minipage}{0.48\textwidth}
        \begin{align}
      \lambda_1=-\frac{\lambda_{12}}{2} \frac{\braket{h_2}^2}{\braket{h_1}^2}\\
  \lambda_2=-\frac{\lambda_{12}}{2} \frac{\braket{h_1}^2}{\braket{h_2}^2}\\
 \lambda_3=-\frac{\lambda_{34}}{2} \frac{\braket{h_4}^2}{\braket{h_3}^2}\\
 \lambda_4=-\frac{\lambda_{34}}{2} \frac{\braket{h_3}^2}{\braket{h_4}^2}
        \end{align}
    \end{minipage}%\bigskip
     \label{2flat}
    \end{subequations}\\
Here, $\braket{h_i}$ is defined in terms of the radial field $\rho$ ($\eta$) and the flat directions $\bf n$ ($\bf l$) as follows
\begin{equation}
\begin{aligned}
    \braket{h_1} &= \braket{\rho} n_1, \\
    \braket{h_2} &= \braket{\rho} n_2, \\
    \braket{h_3} &= \braket{\eta} l_1, \\
    \braket{h_4} &= \braket{\eta} l_2.
\end{aligned}
\end{equation}
The tree-level potential, expressed in terms of the radial fields along the arbitrary directions ${\bf N}$ and ${\bf L}$ as explained above, is given by
\begin{equation}\label{v4scal}
\begin{split}
V_\text{tr}= & \rho^4 \left( \frac{1}{4} \lambda_1 N_1^4+\frac{1}{4}\lambda_2 N_2^4+\frac{1}{4} \lambda_{12} N_1^2 N_2^2\right)\\
&+\eta^4 \left( \frac{1}{4}\lambda_3 L_1^4+\frac{1}{4}\lambda_4 L_2^4+ \frac{1}{4} \lambda_{34} L_1^2 L_2^2 \right)\\
 &+\rho^2 \eta^2  \bigg( \lambda_{0} N_1 N_2 L_1 L_2+\frac{1}{4} \lambda_{13} N_1^2 N_1^2+\frac{1}{4} \lambda_{14} N_1^2 L_2^2\\
 &+\frac{1}{4} \lambda_{23} N_2^2 L_1^2+\frac{1}{4} \lambda_{24} N_2^2 L_2^2 \bigg).
\end{split}
\end{equation}
The solutions for the flat directions ${\bf n}$ and ${\bf l}$ \footnote{Two flat directions ${\bf n}$ and ${\bf l}$ may be interpreted as a {\it flat plane.}} that satisfy Eq. (\ref{2flat}) which itself meets the conditions of Eq. (\ref{vcon}) for the tree-level potential in Eq. (\ref{v4scal}), are
\begin{equation}
\begin{split}
 n_1^2=\frac{\lambda_{12}}{\lambda_{12}-2\lambda_{1}},~ n_2^2=\frac{2\lambda_1}{2\lambda_{1}
 -\lambda_{12}},\\
 l_1^2=\frac{\lambda_{34}}{\lambda_{34}-2\lambda_{3}},~ l_2^2=\frac{2\lambda_3}{2\lambda_{3}-\lambda_{34}}.
 \end{split}
\end{equation}
Considering the aforementioned flatness conditions, the requirement for the potential to be bounded from below is simplified to $\lambda_0<0,~ \lambda_{12}<0$ and $\lambda_{34}<0$. Alternatively, the flatness condition can be expressed as
\begin{equation}
\begin{split}
\lambda_{12}^2=4 \lambda_1 \lambda_2,\\
\lambda_{34}^2 =4 \lambda_3 \lambda_4,\\
  \lambda_{13}\lambda_{24}=\lambda_{14}\lambda_{23}.
 \end{split}
\end{equation} 
Now, It can be shown that the eigenvalues of the mass matrix are
\begin{equation}\label{masses}
\begin{split}
&m_1^2=0, \\
&m_2^2=0 ,\\
 &m_3^2= - \left(\braket{h_1}^2 +\braket{h_2}^2 \right) \left( \frac{\braket{h_3}\braket{ h_4}}{\braket{h_1} \braket{h_2}}\lambda_{0} +\lambda_{12} \right), \\
 &m_4^2= -  \left(\braket{h_3}^2 +\braket{h_4}^2 \right) \left( \frac{\braket{h_1} \braket{h_2}}{\braket{h_3} \braket{h_4}} \lambda_{0} +\lambda_{34} \right).
 \end{split}
\end{equation}
As we asserted earlier, it is clear that there exist two classically massless states along the flat directions, while two other scalars are heavy. 
Radiative corrections should give small mass corrections to both scalons. 

\section{Effective potential}

In general, the radial fields acquire non-zero VEVs, $\braket{\rho}$ and $\braket{\phi}$, introducing two distinct scales into the theory. The calculation of the effective potential, therefore, differs from the GW approach, which involves only one radial field, and consequently, only one scale.
When dealing with the renormalization of a theory with different scales, caution is necessary concerning perturbativity, as these scales may be significantly separated. The intricacies of multi-scale renormalization were elucidated by Einhorn and Jones in \cite{Einhorn:1983fc} (refer also to \cite{Ford:1996yc} for multi-scale renormalization involving two fields). To streamline the computations, we assume that the two scales in our model consistently remain within the perturbativity regime. Therefore, we can express the one-loop effective potential in terms of only one scale $\Lambda$
\begin{equation}
\begin{split}
V_\text{eff}^\text{1-loop}(\rho {\bf n}, \eta {\bf l})=A({\bf n})\rho^4+B({\bf n})\rho^4 \log \frac{\rho^2}{\Lambda^2}\\
+A'({\bf l})\eta^4+B'({\bf l})\eta^4 \log \frac{\eta^2}{\Lambda^2}\,.
\end{split}
\end{equation}
where the coefficients are given by
\begin{equation}
\begin{split}
& A({\bf n})=\frac{1}{64 \pi^2 \braket{\rho^4}}\sum_{i=3,4} m_i^4 \left(-\frac{3}{2}+\log \frac{m_i^2}{\braket{\rho}^2}\right) \\
& B({\bf n})= \frac{1}{64 \pi^2 \braket{\rho^4}}\sum_{i=3,4} m_i^4\\
& A'({\bf l})=\frac{1}{64 \pi^2 \braket{\eta^4}}\sum_{i=3,4} m_i^4 \left(-\frac{3}{2}+\log \frac{m_i^2}{\braket{\eta}^2}\right) \\
& B'({\bf l})=\frac{1}{64 \pi^2 \braket{\eta^4}}\sum_{i=3,4} m_i^4\\
\end{split}
\end{equation}
The one-loop effective potential needs to fulfill the tadpole conditions along two directions,
\begin{equation}
\frac{\partial V_\text{eff}^\text{1-loop}}{\partial \rho} \bigg|_{\rho=\braket{\rho}} =\frac{\partial V_\text{eff}^\text{1-loop}}{\partial \eta} \bigg|_{\eta=\braket{\eta}}=0.
\end{equation} 
These conditions determine the scale $\Lambda$ as
\begin{equation}
\Lambda=\braket{\rho} \exp\left(\frac{A}{2B}+\frac{1}{4}\right)=\braket{\eta} \exp\left(\frac{A'}{2B'}+\frac{1}{4}\right),
\end{equation}
leading to the following effective potential 
\begin{equation}
\begin{split}
V_\text{eff}^\text{1-loop}=B({\bf n})\rho^4 \left( \log \frac{\rho^2}{\braket{\rho}^2}-\frac{1}{2}\right)\\+B'({\bf l})\eta^4\left(  \log \frac{\eta^2}{\braket{\eta}^2}-\frac{1}{2}\right).
\end{split}
\end{equation}
Following the procedures outlined in \cite{Alexander-Nunneley:2010tyr,Ghorbani:2015xvz}, one can determine the masses of the two scalons resulting from the above effective potential,
\begin{equation}
\delta m_1^2=\frac{1}{8\pi^2 \braket{\rho}^2}\sum_{i=3,4} m_i^4,~~~~~
\delta m_2^2=\frac{1}{8\pi^2 \braket{\eta}^2}\sum_{i=3,4} m_i^4,
\end{equation}
where $m_3^2$ and $m_4^2$ are given by Eq. (\ref{masses}).

\section{Conclusion}
It has been well-accepted that in scale invariant multi-scalar theories, according to Gildener-Weinberg assumption \cite{Gildener:1976ih}, along a flat direction there exist only one classically massless scalar due to scale symmetry breaking dubbed as scalon, while all other scalars are massive. In this paper, we have proposed a toy model with four real singlet scalars that possesses two classically massless states. This model can be extracted from more phenomenological models such as two Higgs doublet model in addition to two real singlet scalars, but our staring point is a hypothetical potential with no internal or additional global symmetries. 
Different from the Gildener-Weinberg scheme, we have constructed two flat directions which leads to two classically massless states or two scalons. Note that none of the massless states are due to space-time independent symmetries like global $U(1)$ symmetry. The key point is that one is not obliged to define only one flat direction as in GW approach. We have also calculated the masses of massive states at tree level, and presented the one-loop effective potential and radiative corrections to the masses of two scalons.  Incorporating more than two scalons in a renormalizable theory within the framework of the generic GW potential seems not feasible.

\bibliographystyle{plain}
\bibliography{ref.bib}

\end{document}